\documentclass[useAMS,usenatbib]{mn2e}
\usepackage{graphics}

\def\nii{[N~{\sc ii}]}      
\def\oiii{[O~{\sc iii}]} 
      
\def\ha{H$\alpha$}          
\def\stry{{Str\"omgren {\it y}}}    
\def\rr{{$r^\prime$}}     
\def\sexa{{Sextans~A}}    
        
\def\sexb{{Sextans~B}}  
\def\wlm{{WLM}}  
\def\gr8{{GR~8}}  
\def\hii{H~{\sc ii}} 
\def\hi{H~{\sc i}}               
\def\arcdeg{\hbox{$^\circ$}}   
\def\arcmin{\hbox{$^\prime$}}   
\def\arcsec{\hbox{$^{\prime\prime}$}}        
\def\gg{{$g^\prime$}}

\title[Searching for PNe in IC~1613, WLM and GR8]{The Local Group
Census: searching for PNe in IC~1613, WLM and GR8\thanks {Based on
observations obtained at the 2.5m~INT telescope (La Palma, Spain) operated 
by the Isaac Newton Group and at the 8.2m~VLT telescope (Paranal, Chile)
operated by ESO (Proposal 70.B-0580(A)).}}

\author[L. Magrini et al.]{L. Magrini$^{1}$\thanks{E-mail:
laura@arcetri.astro.it}, R.L.M. Corradi$^{2}$, R. Greimel$^{2}$, 
P. Leisy$^{2}$, A.     Mampaso$^{3}$,
 \newauthor    
 M. Perinotto$^{1}$,  J.R.   Walsh$^{4}$,  N. A. Walton$^{5}$, A.A.   Zijlstra$^{6}$, D. Minniti$^{7}$, and M. Mora$^{8}$.\\ 
$^{1}$Dipartimento di Astronomia e Scienza dello Spazio, Universit\'a di     
Firenze, L.go E. Fermi 2, 50125 Firenze, Italy \\    
$^{2}$Isaac Newton Group of Telescopes, Apartado de Correos 321, 38700 Santa       
Cruz de La Palma, Canarias, Spain \\   
$^{3}$Instituto de Astrof\'{\i}sica de Canarias, c. V\'{\i}a L\'actea s/n,      
38200, La Laguna, Tenerife, Canarias, Spain\\      
$^{4}$ST-ECF, E.S.O., Karl-Schwarzschild-Strasse 2, 85748 Garching bei M\"{u}nchen, Germany\\         
$^{5}$Institute of Astronomy, University of Cambridge, Madingley Road, Cambridge CB3 0HA, UK \\
$^{6}$Physics Department, UMIST, P.O. Box 88, Manchester M60 1QD, UK \\ 
$^{7}$Department of Astronomy, Cat\'olica de Chile, Pontificia Universidad, Vicuna Mackenna 4860, Casilla 306, Santiago 22\\
$^{8}$E.S.O., Karl-Schwarzschild-Strasse 2, 85748 Garching bei M\"{u}nchen, Germany\\     
}

\begin{document}

\date{}

\pagerange{\pageref{firstpage}--\pageref{lastpage}} \pubyear{2004}

\maketitle

\label{firstpage}

\begin{abstract}
In the framework of the Local Group Census (LGC),
a survey of the Local Group (LG) galaxies above Dec=$-30^\circ$,  
which is aimed at surveying their populations with strong emission
lines, we have
searched for planetary nebulae (PNe) in the low-metallicity dwarf
irregular galaxies IC~1613, \wlm, \gr8. 
Two new candidate PNe have been found in IC~1613, one in WLM and none in \gr8.
The  observations presented in this paper, together with the previous results from the LGC,
represent the first step in the study of the PN population in 
low-metallicity, dwarf irregular galaxies of the Local Group.  
They will be followed by deep spectroscopy to confirm 
their nature and to study their physical-chemical properties.
We used the observed number of PNe in each LG galaxy to estimate a lower limit to the mass of the
intermediate-age population  which was compared  to the
Star Formation Rate (SFR) of LG dwarf galaxies.
These results are in agreement with those from accurate star formation history (SFH) 
analysis for these small galaxy systems. 
\end{abstract}

\begin{keywords}
planetary nebulae: individual : IC~1613, WLM--    
Galaxies: individual: IC~1613, WLM, GR~8
\end{keywords}

\section{Introduction}
\label{Sect-intro}

The Local Group consists of three spiral galaxies and a large
number of dwarf galaxies ($\sim$90\% of its 40 known members).
This proportion of galaxy types may be typical of the local Universe as
similar distributions are known to exist in nearby groups
\citep{miller} and clusters \citep{phillips}.  The dwarf galaxies
(irregular, spheroidal or elliptical) of the LG are of particular
interest as their proximity allows us to study them in detail, testing
predictions concerning their formation and evolution,  as, for
instance, the evolution of different kind of dwarfs \citep{richer95} or
their star formation histories \citep{aparicio01}.

Furthermore, their study is relevant because, according to the
hierarchical scenario of galaxy formation, dwarf galaxies are the
first structures to form and from their merging larger
galaxies are built \citep{nagashima}.  Moreover, low-luminosity dwarf
galaxies, as those discussed in this paper, are metal poor and are
expected to have abundances close to primordial values 
\citep{kunth}. 
Thus the abundances derived there can be useful in
extrapolating primordial He/H (\citealt{peimbert74}; \citealt{lequeux79}).

It is also interesting to compare the overall star formation history
(SFH) of the Universe predicted by \citet{madau} with the SFHs of
galaxies of different morphological type that exist in the LG 
(\citealt{mateo}, \citealt{aparicio01}). 
This is fundamental in order to understand 
the variation of SFH from the global average in galaxies of differing  
morphological type (\citealt{renzini95}, \citealt{mcgaugh94}).

A quantitative SFH can be obtained with the colour-magnitude diagram
(CMD) technique (cf. \citealt{aparicio04}). 
This technique involves three main ingredients: i) the data,
from which a deep observational CMD can be plotted; ii) stellar
evolution and a bolometric correction libraries providing colors and
magnitudes of stars as a function of age, mass, and metallicity; and
iii) a method to relate the number of stars populating different regions of
the observational CMD with the density distribution of stars 
as a function of age, mass, and metallicity, as predicted by 
stellar evolution theory (see \citealt{aparicio04} for a complete
review).  This method can, in principle, constrain the entire star
formation and chemical history of a galaxy, but, of course, it depends
on the precision of the input models and on the assumptions \citep{aparicio04}.  
Moreover, since the overall results of this
method are model dependent, complementary specific evolutionary phases
can be used as {\em age-tracers} to mark different populations and to
infer the presence of a given age component \citep{mateo}.

In this paper, we present new results from the latest study of individual group 
members as part of the continuing LG census. 
This census is a narrow- and broad-band imaging survey of the galaxies of the LG, 
visible from the Isaac Newton Group of telescopes (ING),  whose aim is to find, 
catalogue and study old and 
young emission-line populations, such as planetary nebulae, supernova
remnants and \hii\ regions \citep{corradi04B}. We are
interested in identifying Planetary Nebulae, which are good indicators
of the presence of intermediate-age stars. 
LGC observations were obtained with the Wide Field Camera (WFC) at the 2.5m 
Isaac Newton telescope (INT). The detector of the WFC is a mosaic composed of 
four CCDs, with 2048$\times$4096 pixels each. It covers a field of view of $34\arcmin
\times 34\arcmin$.  The pixel scale is 0.$\arcsec$33. \footnote{See  
{\tt http://www.ing.iac.es/$\sim$rcorradi/LGC} for the description of the
project}

The results obtained until now 
in the search for PNe in the LG galaxies are described in the following papers: \citet{M02},
\citet{M03}, hereafter M03; \citet{corradi05}; \citet{leisy};
\citet{corradi04B}.

Here we describe the results of the observations of the next
three targets: the dwarf irregular galaxies IC~1613 (morphological
type Ir~V according to \citet{vdb00}, hereafter vdB00), \wlm\
(Ir~IV-V; \citealt{vdb00}), and \gr8\ (dIrr;  \citealt{mateo}).
 
The gas-rich dwarf irregular galaxy IC~1613 was first discovered by
\citet{wolf06} and because of its proximity (730 kpc,
\citealt{dolphin01}), its high Galactic latitude (-60\arcdeg.6) and
consequently small Galactic extinction, has provided an excellent
opportunity to observe stellar populations in a low metallicity
environment. In fact, its oxygen abundance as derived from \hii\
regions is approximately 12 + log (O/H)=7.7 (\citealt{skillman};
\citealt{lee03}).  Moreover, IC~1613 is a relatively isolated
non-interacting irregular galaxy, as proven by the lack of stellar
clusters generated by interactions with other galaxies \citep{vdb00B}.
From an evolutionary point of view, it is a very primitive galaxy
because the stellar mass is comparable to the gas mass
\citep{hodge91}.  \citet{freedman} presented B, V, R and I stellar
photometry pointing out the presence of stars in a wide range of ages,
from young to intermediate and old population.  Recent Hubble Space
Telescope observations by \citet{skillman03} found an enhanced star
formation rate (SFR) from 3 to 6 Gry ago.  A survey for emission-line
objects was done in this galaxy by \citet{lequeux} who found one
candidate PN.

The Wolf-Lundmark-Melotte (\wlm) galaxy was discovered by
\citet{wolf23} and independently by \citet{lundmark} and by
\citet{melotte}.  \wlm\ (called also DDO~221, \citealt{vdb66}) is a Local Group
metal-poor dwarf irregular (12 + log(O/H) = 7.7; \citealt{skillman}),
in a part of the Local Group relatively free of galaxies, at a
distance of 925~kpc (vdB00).
\citet{dolphin00} studied with the Hubble Space Telescope the SFH in a
portion of WLM.  It appears to have begun no more than 12 Gyr ago,
forming more than half of its stellar population by 9 Gyr ago.  The SFR
has subsequently gradually decreased until a recent increase in activity
starting between 1 and 2.5 Gyr ago, but is still continuing to the
present time.  Recent star formation is indicated by  several \hii\ regions 
around O-type stars, as observed by \citet{hodgemiller}.

To date no candidate PN is known in this galaxy.   
In fact \citet{minniti} found the two 
candidate PNe  proposed by \citet{jacoby}  to be ordinary stars.
The other candidate PN identified by Minniti \& Zijlstra (1996;1997)
was found to be a compact \hii\ region (Zijlstra, private communication).

\gr8  was first discovered by  \citet{reaves56} and it was catalogued as 
DDO 155 by \citet{vandenbergh59}.
It is a suspected member of the Local Group (vdB00), but
observations of a single Cepheid by \citet{tolstoy} and of the tip of
the RGB by \citet{dohm} placed this galaxy at the distance of 2.2~Mpc,
beyond the usually accepted limit of the LG.  \gr8\ is one
of the intrinsically smallest irregular galaxies known, with an
effective radius of only 50~pc and \hi\ mass of 10$^6 M_{\odot}$
\citep{hodge89}.  Several \hii\ regions are known in this galaxy
\citep{hodge74}. Their abundances have been measured by
\citet{skillman88}, finding one of the lowest known oxygen abundances
(12 + log (O/H)=7.4).  No PN have been identified in \gr8\
\citep{jacoby}.
Basic data on the three galaxies are reported in Table~\ref{Tab_gal}.

\begin{table*}    

\caption{Data on the observed  galaxies.} 
 
\begin{center} 
\begin{tabular}{l l r l l l l c l}     
\hline     
Name  & \multicolumn{2}{c}{R.A. (2000.0) Dec.} & Type &  Distance  & Optical size & M$_V$ & 12+$\log$(O/H) & M  \\  
      & 	   &			       &      & (kpc)	   &		  &	  & ~~$\dag$	   &(M$_{\odot}$)\\
\hline  
IC~1613 & 01$^h$ 04$^m$ 47$^s$.8 &  +02\arcdeg\ 07\arcmin\ 04\arcsec & Ir~V~(1)	& 725~(1)  & 16\arcmin$\times$20\arcmin  & -15.3~(1) &7.7~(7) & 1$\times$10$^{8}$~(10) \\
      & 	   &			       &      & 	   &3.4~kpc$\times$4.2~kpc~(4)&	  &		 &\\
WLM     & 00$^h$ 01$^m$ 58$^s$.1 &  -15\arcdeg\ 27\arcmin\ 39\arcsec & IrIV-V~(1)   & 925~(1)  & 6.5\arcmin$\times$12.6\arcmin  & -14.4~(1) & 7.7~(8) & 1.5$\times$10$^{8}$~(2) \\
      & 	   &			       &      & 	   &1.7~kpc$\times$3.4~kpc$_{(5)}$&	  &		 &\\
GR~8    & 12$^h$ 58$^m$ 07$^s$.4&  +14\arcdeg\ 13\arcmin\ 03\arcsec & dIrr~(2)	& 2200~(3) & 1.1\arcmin$\times$1.0\arcmin	& -11.6~(2) &7.4~(9) & 7.6$\times$10$^{6}$~(1) \\
      & 	   &			       &      & 	   &0.7~kpc$\times$0.6~kpc~(6)&	  &		 &\\
\hline
\end{tabular}
\label{Tab_gal} 
\end{center} 
References: R.A. and Dec. come from NED. 
(1) \citet{vdb00}; (2) \citet{mateo}; (3) \citet{tolstoy};
(4) \citet{ables}; (5) \citet{gallouet75}; (6) NED; 
(7) average of values obtained by \citet{skillman} and \citet{lee03}; (8) \citet{skillman}; (9)  \citet{skillman88};
(10) \citet{lake}. 
$\dag$ Oxygen abundances reported in the Table have been computed from 
spectra of \hii\ regions. 
\end{table*}

\section{Observations} 

IC~1613 (01$^h$ 04$^m$ 47$^s$.8 +02\arcdeg\ 07\arcmin\ 04\arcsec,
J2000.0), \wlm\, (00$^h$ 01$^m$ 58$^s$.1 -15\arcdeg\ 27\arcmin\
39\arcsec, J2000.0), and \gr8\ (12$^h$ 58$^m$ 07$^s$.4 +14\arcdeg\
13\arcmin\ 03\arcsec, J2000.0) were observed with the WFC at the prime focus 
of the INT , on February 2001, October 2002, and September 2004 (see Tab.\ref{Tab_gal}).

The three galaxies were observed with the following filters: \oiii\
(500.8/10.0~nm), \ha + \nii\ (656.8/9.5), \stry\ (550.5/24.0), \rr\
(Sloan r, 624.0/135), and \gg\ (Sloan g, 484.6/128).  We used the
\stry\ and the \gg\ filters as off-band images for the continuum
subtraction of the \oiii\ images, while  \rr\ was used as off-band 
for \ha+\nii.  The February 2001 nights in which we observed \gr8,
were photometric, whereas the October 2002 nights were not. The night
of September 2004, which was photometric, was used to calibrate the
observations of October 2002.  Each exposure was split into several
sub-exposures.  The total exposure times,   the number of exposures, and the seeing  in each filter 
are listed in Tab.\ref{Tab_obs}.

\begin{table}    

\caption{Summary of the observations: target galaxies, date of
observations, telescope and instrument, total exposure times,  number of 
exposures, seeing, and filters.  }
\begin{center} {\scriptsize
\begin{tabular}{l l l l l l l}     
\hline  
Target  & Date     & Inst.   & Exp.  & N.& Seeing     & Filter \\  
        &          &         & (sec) &   &(\arcsec) &  \\  
\hline\hline    
Gr8    	&Feb. 2001 & WFC@INT &4800 & 4	&1.0	&\ha+\nii\ \\
	&	   & 	     &3600 & 3	&1.0	& \oiii\	  \\
	&	   & 	     &1800 & 3	&1.0	& Str. y   \\
	&	   & 	     &2400 & 5	&1.0	& \rr	  \\	
	&	   & 	     &2400 & 5	&1.0	& \gg	  \\
\hline\hline
IC1613 &Oct. 2002  & WFC@INT & 3600 & 3	&1.3 &\ha+\nii \\
	&	   & 	     & 4800 & 4	&1.3 & \oiii	  \\
	&	   & 	     & 1800 & 4	&1.4 & Str. y   \\
	&	   &	     & 1200 & 3	&1.3 & \rr	  \\	
	&	   &	     & 1200 & 3	&1.7 & \gg	  \\	
	&Sept. 2004& 	     & 300  & 1	&1.1 &\ha+\nii \\
	&	   &	     & 300  & 1	&1.2 & \oiii	  \\
\hline
IC1613	&Sept. 2004 &FORS2@VLT& 200 & 2	&0.9 & \ha \\
	&	   &	      & 20  & 2	&0.9 & R  \\ 
\hline\hline
WLM	&Oct. 2002 & WFC@INT &5600 & 6	&1.3	&\ha+\nii \\
	&	   & 	     &3200 & 6	&1.6	& \oiii	  \\
	&	   & 	     &1600 & 4	&1.6	& Str. y  \\
	&	   & 	     &4000 & 11	&1.2	& \rr	  \\	
	&	   & 	     &1800 & 4	&1.6	& \gg	  \\
\hline
WLM	&Oct. 2002 & FORS1@VLT&300 &	&0.8 &\oiii \\
	&	   &	     & 300 &	&0.8	 &\oiii/6000 \\
\hline
\end{tabular}
}    
\end{center} 
\label{Tab_obs}
\end{table}

Several observations of the spectrophotometric standard stars
BD+33~2642 and G191-B2B \citep{oke} were made each night during the
February 2001 run, while Feige 110 \citep{oke} was observed during the
September 2004 run.  

We complemented our observations with images of
\wlm\ from the ESO archive {\footnote{\tt http://archive.eso.org}, taken with
VLT+FORS1 on October 2002, and with observations of IC~1613 
we obtained on September 2004 with VLT+FORS2 (see Tab.\ref{Tab_obs}).  

The filters used  for the FORS1 observations  were
\oiii\ (500.5/0.8 nm) and \oiii/6000 (510.9/0.8 nm),  while for 
FORS2 observations \ha\ (656.3/6.1~nm) and  R (655.0/165.0~nm)}
were employed. \oiii/6000  and  R filter images were used for continuum subtraction 
of the  \oiii\ and \ha\ images, respectively.

The VLT data taken with the \oiii\ filter  were calibrated
using the INT observations.  Scaling the diameters of INT and VLT
telescopes with the exposure time, the INT images should be as deep
as the VLT images. 
Because of the better seeing the VLT images through the \oiii\ filter  are however about 1 mag deeper 
than the corresponding \oiii\ INT images.

\section{Data reduction and analysis}     
 
The data reduction was done using IRAF\footnote{IRAF is distributed by the
National Optical Astronomy Observatories, which is operated by the
Association of Universities for Research in Astronomy, Inc.  (AURA)
under cooperative agreement with the National Science Foundation}.
First the data were processed by the ING WFC data-reduction pipeline
\citep{irwin}: they were de-biased, flat-fielded, and
linearity-corrected.  Then they were corrected for geometrical
distortion and subsequently all frames were aligned to the \oiii\ one.
The images were combined taking their median when the number of exposures 
was larger than three, otherwise the average was preferred. 
The numbers of images per filter and per field are specified in Table~\ref{Tab_obs}. 
The cosmic rays were
removed with the CRREJECT algorithm, and finally the sky background
was subtracted.

In order to search for emission-line objects, in  \gr8\ we subtracted
from the \oiii\ frames the properly scaled \stry\ frames, whereas for
IC~1613 and \wlm\ we used the \gg\ frames as a continuum. 
For each galaxy we have chosen the best quality continuum, \stry\ or \gg, to 
do the off-band subtraction.

For the VLT images of \wlm, the scaled  \oiii/6000 images were subtracted 
from the  \oiii\ images.
For the \ha+\nii\ images of all
galaxies, we used the correctly  scaled \rr\ frames as a continuum. 
Finally for the \ha\ images  of IC~1613 taken with VLT, R was used as a 
continuum.
The search for unresolved emission-line objects in the 
continuum subtracted images was done visually at least three times 
per galaxy by different members of the team.  
In addition, we performed photometry of the unresolved sources in 
all the INT images with DAOPHOT \citep{stetson87} to build \ha+\nii - \rr\  
vs \oiii\ - \gg\ diagrams, obtaining in IC~1613 the same results 
as the visual search, while the PN of WLM was not recovered 
because of its proximity to a star (see Fig.\ref{Fig_high}).  

The astrometric solutions were computed using the IRAF tasks CCMAP and
CCTRAN and the APM POSS1 \footnote{\tt
http://www.ast.cam.ac.uk/$\sim$apmcat/} and USNO A2.0 
\citep{monet} catalogues. The final accuracy was $\sim$0\farcs3 {\it r.m.s.}.

\section{Candidate planetary nebulae} 
     
We searched for PNe in our continuum-subtracted frames selecting
objects which satisfy the following criteria \citep{M00}: {\em i)} they must 
appear in both the \oiii\ and \ha+\nii\ images, but not in
the continuum frames; and {\em ii)} they must be unresolved at the distance of
IC~1613, \wlm, and \gr8, on account of the typical physical size of PNe
(0.1-1~pc).  We found  two objects in IC~1613 fulfilling these
criteria, one in \wlm\ and none in \gr8.  Using both \oiii\ and
\ha+\nii\ images, the possibility that these PNe are background
emission line galaxies \citep{mendez93} can be excluded.

\subsection{The completeness limit} 
The incompleteness in searching for emission-line objects 
results from a combination of the probability of missing an
object in the emission-line image and of the probability of wrongly
identifying a star in the continuum frame.
To estimate the completeness limit of our search for PNe, i.e. the magnitude corresponding 
to a probability of non-identification of an emission-line object larger than 50\% 
(defining the incompleteness according to \citealt{minniti}), 
we added  `artificial stars' with various 
m$_{\rm[O~III]}$  within the
range of luminosities expected for PNe \citep{jacoby89}, 
in both \oiii\  and continuum images. 
We have then computed the recovery rate of such artificial objects (\citealt{M02}, M03).
We find that in the three galaxies  the recovery rate is  about 50\%\
for objects with 25.5$\le$$m_{\rm [O~III]}$$\le$26.5
in GR~8, and 24.0$\le$$m_{\rm [O~III]}$$\le$25.0 in
IC~1613 and WLM using the INT images, while in the VLT images for objects with 25.0$\le$$m_{\rm [O~III]}$$\le$26.0.  
The recovery rate is 100\%\ for brighter objects.
The difference in the incompleteness magnitude listed above comes
mainly from the difference in seeing of the data used.
We noted that the three candidate PNe are placed at the edges of each system 
and not in their more populated regions. This might be due to the
identification technique, which is less effective in very crowded regions.  

\subsection{The expected number of PNe}

The number of PNe, as well as the number of stars in any post-main-sequence phase, 
is proportional to the luminosity of the host galaxy, as derived from 
a  simple (i.e. coeval and chemically homogeneous) stellar population model 
\citep{renzini}.
Thus, a small number of PNe is expected in low-luminosity galaxies, such as 
the ones surveyed (see also  Fig.~2 of M03). 
This trend is  also expected  when the galaxies  have, in addition, a  low-metallicity; 
since a fall in the observed number of PNe has been suggested when [Fe/H]$<<-$1.0 (see Fig.~3, M03).

As described above, IC~1613, \wlm, and \gr8\ are low-luminosity and low-metallicity galaxies, so
their expected PN number is very small.  From the V-band luminosity vs. number of 
PNe diagram (Fig.~2 of M03), 
and assuming a search as deep as that
made for Sex B, i.e. complete up to mag$_{\rm [OIII]}$=24.5 (M03), 
we infer that the expected observable number of PN for IC 1613 is
$\sim$10 and $\sim$3 for WLM; while none is expected in GR8.

\subsection{GR~8 and WLM}
The result in \gr8\ was expected from the low- luminosity of
this galaxy and is in agreement with the survey of \citet{jacoby}
which did not find any PN.  
The result for \wlm\ also statistically
agrees with the small expected PN population size and the large
distance to this galaxy.  
The PN found in \wlm\ is barely detectable
in LGC images, whereas it is very clear in the VLT FORS1 images  (see Fig.~\ref{Fig_high}). 
Its position and fluxes are  reported in Table~\ref{Tab_pos}. It is  marked in
Figure~\ref{Fig_pos_wlm}.  In Fig.~\ref{Fig_high}
thumbnail images taken in the various filters 
of the PN are shown. It can be seen in the VLT \oiii\ images  that the PN is projected very 
close to a star, which is the object near to the location 
of the PN in the continuum image and  disappears in the continuum-subtracted images.   
In addition, we found that the  two objects
previously claimed to be candidate PNe \citep{jacoby} are instead
stars because of their strong continuum emission in all broad band images, 
as already recognized by \citet{minniti}.

\subsection{IC~1613}
The objects
belonging to IC~1613 presented in this paper are the first candidate
PNe discovered in this galaxy.  \citet{lequeux} in an
objective-prism survey for emission-line objects in this galaxy
identified one candidate PN, but neither a following survey for \hii\
regions \citep{skillman} nor this work could confirm this object.  The
two candidates lie inside the optical size of the galaxy
(16\arcmin$\times$20\arcmin, according to \citealt{ables}).
The position and fluxes of the PN candidates are listed in  Table~\ref{Tab_pos} 
and marked in Figure~\ref{Fig_pos}. In Fig.~\ref{Fig_high}
thumbnail images of these PNe taken in the various filters are shown.

\begin{figure}
\resizebox{\hsize}{!}{\includegraphics{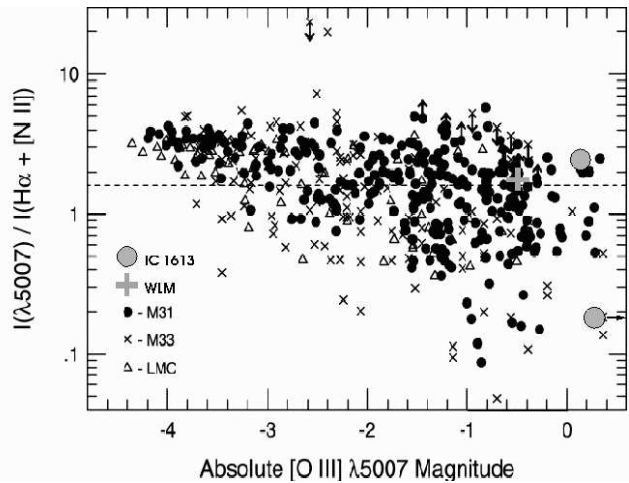}} 
\caption{Absolute magnitude vs. I(\oiii)/I(\ha+\nii) from \citet{ciardullo02}.
The line ratios are for extragalactic PNe. 
In this diagram the location of the three discovered candidate PNe are shown. }
\label{Fig_ciard}
\end{figure}

\begin{figure*}
\resizebox{18.cm}{!}{\includegraphics{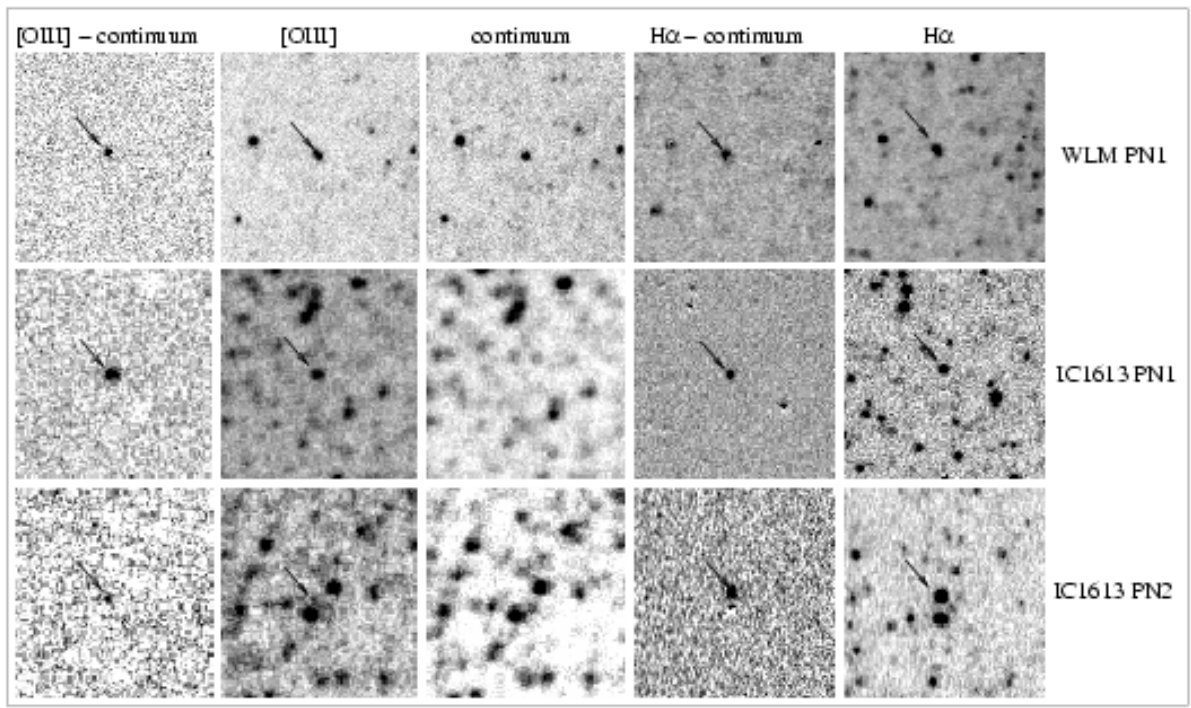}} 
\caption{\oiii-{\it continuum}, \oiii, {\it continuum}, and \ha-{\it continuum} images of the discovered PNe
are shown. The size of each image is approximately  0.7\arcmin$\times$0.7\arcmin.  
For WLM PN1: the \oiii and  \oiii/6000 ({\it continuum}) images are from VLT, and \ha-\rr\ from INT.
For IC~1613 PN1 and PN2: \oiii and \gg\ ({\it continuum}) images are from INT, and \ha-R from VLT.
North is at the top, East to the left.  }
\label{Fig_high}
\end{figure*}

\begin{figure}

\centering  
\resizebox{\hsize}{!}{\includegraphics{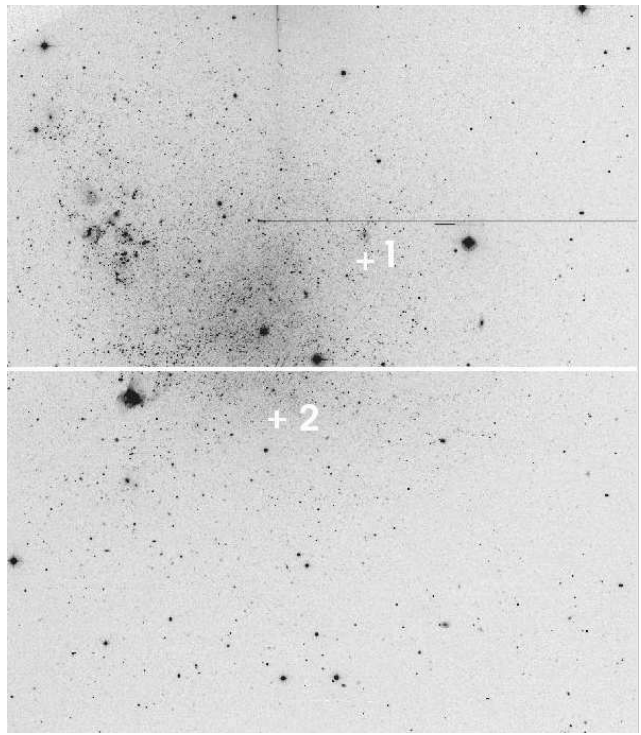} }   
\caption{WFC \oiii\ image of the galaxy IC~1613. The gap is due to the  
physical separation among  the mosaic CCDs.      
North is at the top, East to the left.  
The size  is approximately    15\arcmin$\times$19\arcmin.  
Candidate PNe are marked with a cross. Their identification number is that of   
Table~\ref{Tab_pos}. }
\label{Fig_pos}    
\end{figure}

\begin{figure}
\centering 
\resizebox{8.cm}{!}{\includegraphics{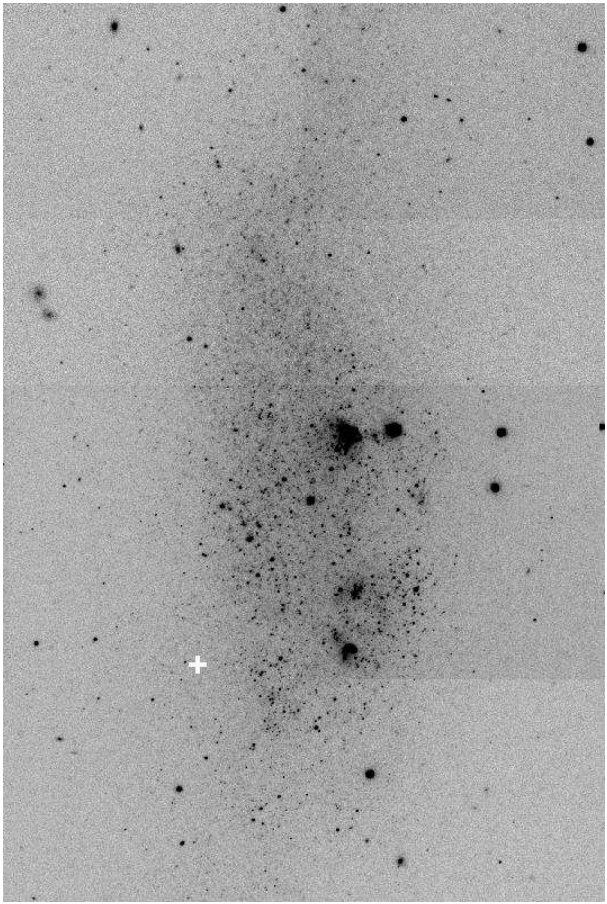}} 
\caption{VLT  \oiii\ image of the galaxy \wlm.  
North is at the top, East to the left.  
The size  is approximately   7\arcmin$\times$12\arcmin.  
The candidate PN is marked with a cross.}
\label{Fig_pos_wlm}
\end{figure}

A criterion which is generally considered when selecting extragalactic PN
candidates is the ratio between the \oiii\ and the \ha\ + \nii\
fluxes, which allows a statistical discrimination with unresolved \hii\
regions.  The criterion adopted by \citet{ciardullo02} is
R=I(\oiii)/I(\ha + \nii)$>$ 1.6 for PNe.  The reason is that central
stars of PNe are generally hotter than the OB stars that excite \hii\
regions, and this produces in PNe an \oiii\ flux higher than the \ha\
+ \nii\ flux. The value 1.6 is an empirical value derived with a large
sample of extragalactic PNe.  The ratio R is a function of the absolute
magnitude of the PN, as we can see in Fig.~\ref{Fig_ciard}, reproduced 
from the paper by \citet{ciardullo02}.

For low luminosity PNe, R ranges from 1.5-2 down to very low values of 
0.1 or less.  As the object designated as PN2 in IC 1613 has an extremely low
absolute magnitude we can accept it as a PN candidate even with its
low R ratio of $\sim$0.2, as can be seen from Fig.~\ref{Fig_ciard}.
In addition, both PN~1 and PN~2 are far from the main star formation regions, 
and this reduces the possibility of contamination with compact \hii\ regions \citep{ciardullo02}. 
Furthermore the value of R might  be effected by the extinction on the observed 
line ratios. The extinction can arise within the Galaxy, along the line of sight to the 
LG galaxy and intrinsic to the galaxy itself. Any extinction would 
depress \oiii\ relative to \ha\ and consequently decrease the value of R. 
For all the galaxies considered in this paper, the Galactic latitudes are very high, 
but localised extinction may still occur. Regarding the galaxies themselves, none are 
gas/dust rich and their internal extinction is very low.    

In spite of the 10 PNe expected, we detected only  two candidate PNe
in this galaxy, which may be a significant difference. 
Anyhow we can never be complete for objects such as PNe - for instance  
in our Galaxy we now know more than 2500 PNe, but 10~000-30~000 are expected \citep{zijlstra91}.  
With reference to Fig.~2 by M03, the relatively large number of PNe found in \sexb\ is
an exception in the context of LG low-luminosity galaxies.  In fact,
we found that \sexa, \wlm, IC~1613 and IC~10 have PNe populations
rather smaller than expected.  M03 discussed the particular case of
the starburst galaxy IC~10 whose central area is covered by large
\hii\ regions thus producing a loss in observable PNe because of
the superposition of the large \hii\ regions.  For
low-metallicity galaxies (like \sexa, \wlm, and IC~1613), M03
suggested that the small number of observed candidate PNe might be
explained by the low metal content of the host galaxies, which would
imply a reduced PN formation.  Any firm conclusion in this sense is
prevented by the small number statistics; deeper surveys would be
needed to improve this situation.

\subsection{The absolute magnitude of the PNe}
Another aspect to be noted is the relative faintness of the discovered
PNe. The absolute magnitude of the cutoff of the PNe \oiii\ luminosity
function (PNLF) in a galaxy with a large population of PNe
is considered constant in large galaxies, with a value M$\star$=$-$4.47
\citep{ciardullo02}. 
For dwarf galaxies, the cut-off magnitude decreases to fainter limits
as a function of the galaxy mass and population size \citep{mendez93}.
In fact, the brightest candidate PNe of IC~1613 and WLM are much
fainter than M$\star$.  Considering the distance modulus of 24.3 and
E(B-V)$\sim$0.03 for IC~1613 (vdB00), we obtain an absolute magnitude
for its brightest PN of $0.1$~mag.  For WLM, where the distance
modulus is 23.8 and E(B-V)$\sim$0.02 (vdB00), we found the absolute
magnitude of the brightest PN to be approximately $-0.5$.  
As said above, this is likely due to the small population size of PNe
in these dwarf galaxies, for which the magnitude of the cutoff,
M$\star$, is shifted towards fainter magnitudes \citep{mendez93}.

\begin{table*}    

\caption{PN candidates in IC~1613 and \wlm. Positions are at J2000.0. 
Observed \oiii500.7 and \ha\ fluxes are given in 10$^{-16}$
erg~cm$^{-2}$~s$^{-1}$. 
}   
\begin{center} 
\begin{tabular}{l r r r r r}     
\hline    
Identification & \multicolumn{2}{c}{R.A. (2000.0) Dec.} & $F_{\rm [OIII]}$ &   
$F_{\rm H\alpha + [NII]}$ & m$_{\rm [O~III]}$  \\  
\hline   
WLM 	PN1 &  0 02 03.33  & -15 29 30.4& 5.6 &3.3 & 24.4\\
\hline    
IC~1613 PN1 &  1 04 32.28 & +02 08 43.5 & 5.2 & 2.3 & 24.5\\  
IC~1613 PN2 &  1 04 43.72 & +02 03 40.7 & 1.5 & 6.7 & 25.8\\   

\hline
\end{tabular}
\label{Tab_pos} 
\end{center} 
\end{table*}    

\section{Intermediate-age star formation history} 

Given the mass range covered by PN progenitors, assumed to be classicaly 
between  0.8M$_{\odot}$ (or a bit higher, 1~M$_{\odot}$, cf. \citealt{phillips01}) 
and 8M$_{\odot}$, the stellar population
from which PNe derive is formed at intermediate ages (roughly 1 to 8
Gyr ago) during the history of a galaxy.  One can evaluate the total
mass of such a stellar population using the theoretical relation
between the total luminosity of the population of PNe progenitors and
the PNe number, obtained under the hypothesis of coeval, chemically
homogeneous stars \citep{renzini}.  The total luminosity of the
population of PNe progenitors, $L_{T_{\rm pPNe}}$, is related to the
number of PNe, $n_{\rm PNe}$, to their lifetime, $t_{\rm PNe}$, and to
the so-called evolutionary flux, $\dot{\xi}_{\rm PNe}$.  The
number of stars with initial mass $0.8M_{\odot}<M<8 M_{\odot}$ per unit luminosity
leaving the Main Sequence each year is then given by the expression
\begin{equation} 
L_{T_{\rm pPNe}}=\frac{n_{\rm PNe}} {t_{\rm PNe} \dot{\xi}_{\rm PNe}}. 
\end{equation}  
Thus, counting planetary nebulae and estimating the mean mass of the
PNe progenitors we have a measure of the total mass of the
intermediate-age stellar population, or at least a lower limit to it,
when the PNe survey is not complete.  Using a lifetime of
$\sim$10~000~yrs for the PN phase, a mean specific evolutionary flux
of 2$\times 10^{-11}$~yr$^{-1}$~$L_\odot^{-1}$ (\citealt{renzini},
\citealt{mendez93}, \citealt{buzzoni}), 
a mean progenitor mass of $\sim$1.5~M$_\odot$ (computed using the 
Initial Mass Function by \citet{scalo98},
corresponding to a luminosity of $\sim$5$L_\odot$), the lower limit 
to the intermediate-age stellar mass in the LG
galaxies can be estimated. The intermediate-age stellar
mass is proportional to the number of PNe, and with the
approximations described above, we found that about 2$\times$10$^6$~M$_{\odot}$
were formed for each PN observed.

We used all the {\em candidate} PNe discovered in the Local Group by several 
authors (see references below) with the 
on-band/off-band or similar techniques, 
and the criteria described in Sect.~4,  considered as {\sl bona fide} PNe 
for the following reasons. 
The first is that  these criteria give a good confidence on the PN nature of the candidate. 
For instance, a follow-up spectroscopic study of PN candidates detected in M~33 by \citet{M00} 
confirmed more than 70\% as PNe \citep{magrini03s}, even if the criterion of R$>$1.6
was not used. Among the remaining 30\% objects, some indetermination remains between 
low-excitation PNe and compact \hii\ regions, which have similar spectroscopic and photometric 
features.   
The second reason is that it is impossible to study the statistical properties of PN population 
in the LG if considering only spectroscopically confirmed PNe since 
spectroscopic observations have been obtained for  only $\sim$10\% of the total number of 
LG PNe. 
In addition, once obtained, spectroscopy cannot provide complete confidence in the determination 
of the nature of the object, in many cases because of the 
faintness of the objects  \citep{jacoby05}. 

\begin{figure}  
\resizebox{9cm}{!}{\includegraphics{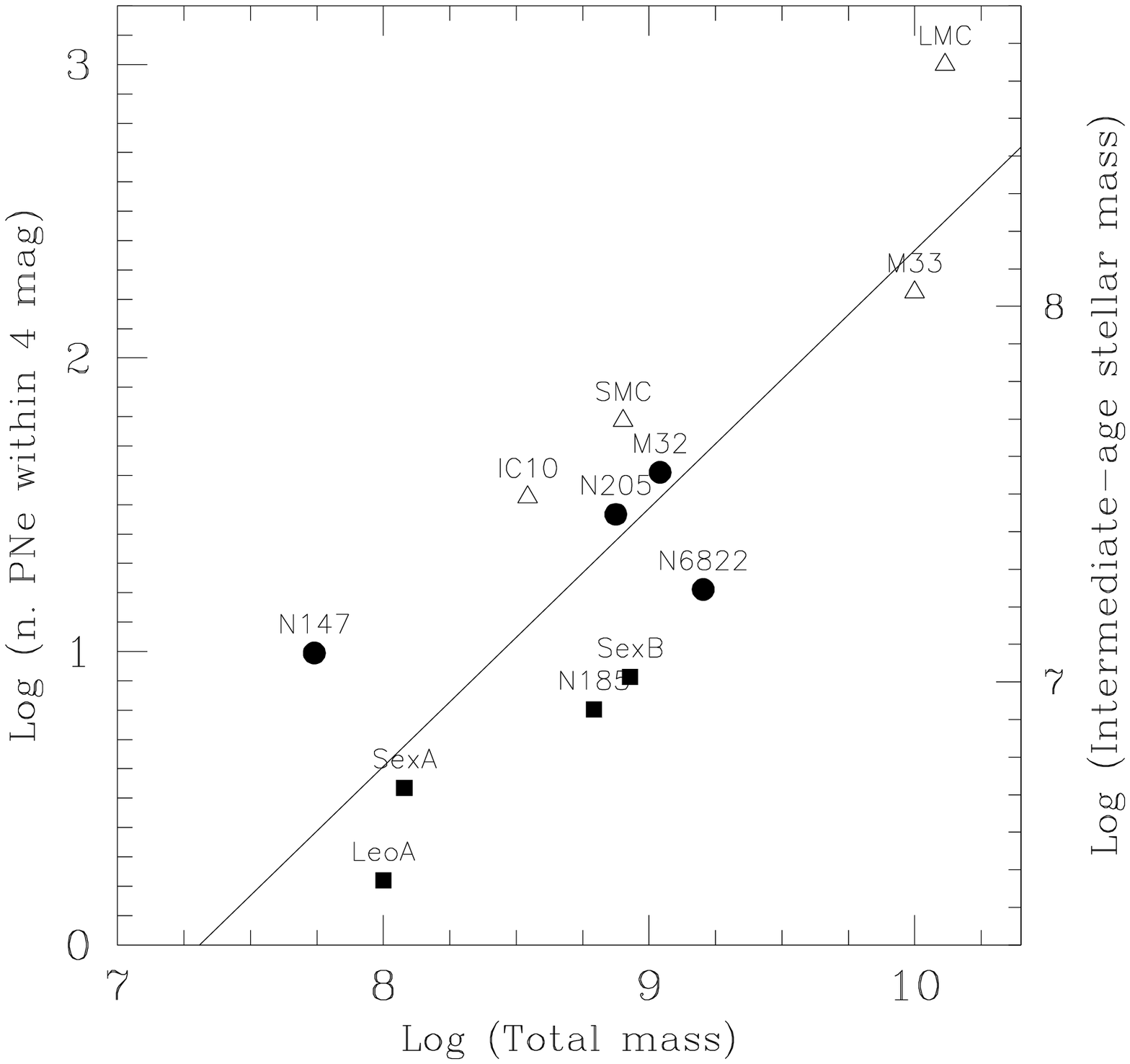}}  
\resizebox{9cm}{!}{\includegraphics{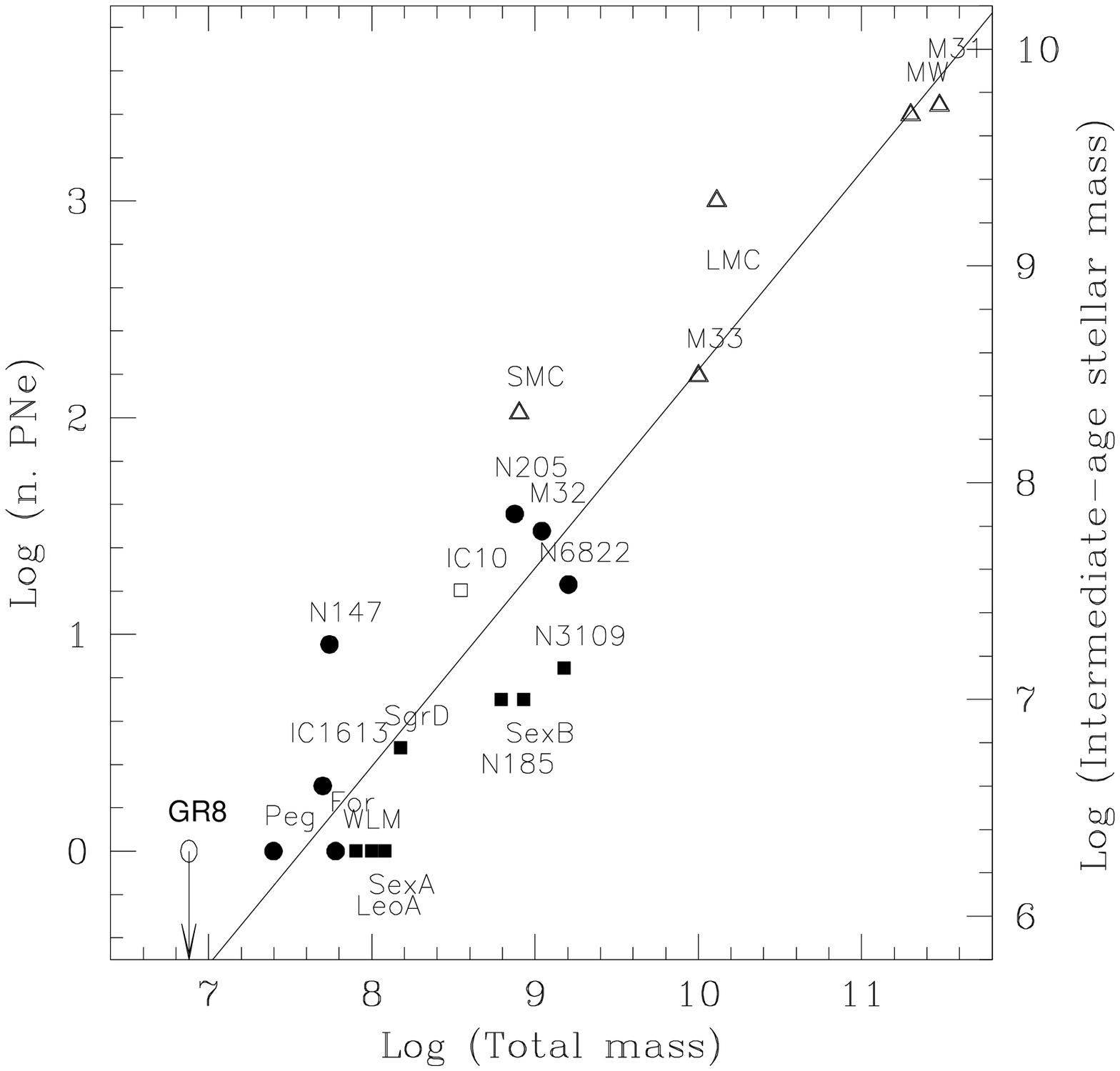}}  
\caption{Upper  panel: the total
stellar mass vs. the number of PNe within 4 mag from the cutoff of the
PNLF in the galaxies where it was possible to extrapolate this value.
Dwarf galaxies are marked with filled circles when SFR$\geq$0.2, and
filled squares when SFR$<$0.2, according to \citep{mateo}. Empty
squares indicate galaxies with insufficient data available to study
their SFHs. Triangles indicate large galaxies.  The continuous line is
the least squares fit. Lower panel: 
the total stellar mass of a galaxy
expressed in M$_{\odot}$ (from vdB00) vs. number of observed PNe (left
vertical axis), which is proportional to the mass of the
intermediate-age stellar population (right vertical axis).}  

\label{Fig-inter}  
\end{figure}

In Figure~\ref{Fig-inter}, we show the result, plotting
the total stellar mass of the galaxies (values from vdB00) vs. the
observed number of PNe (left vertical axis). The latter
is proportional to the the mass of the
intermediate-age stellar population (right vertical axis) with a
constant of proportionality of 2$\times$10$^6$. 
Errors amount to $\sim$30\% in the estimate of the total stellar mass of
galaxies, and to $\sim$50\% in the estimate of the intermediate-age
mass for the galaxies where we obtained the number of PNe within 4~mag 
from the cutoff of the PNLF (upper panel).
For the galaxies where we have only the number of observed PNe (lower 
panel), 
only a lower limit to the intermediate-age mass can be given.
The references on the number of PNe in the Local Group and also on
their spectroscopic confirmation, if any, are: M~31
2764 PNe identified with the Planetary Nebulae Spectrograph by
\citet{merret05}, $\sim$30 confirmed with spectroscopy by
\citet{ciardullo99}, \citet{stasinska98}; MW $\sim$2400 (1143 + 242
true and probable PNe in \citealt{acker92}, and $\sim$1000 in
\citealt{parker03}); M~33 152 PNe identified by \citet{ciardullo04}
($\sim$30 confirmed by \citet{magrini03s}); LMC $\sim$1000 estimated, 
350 discovered by \citet{jacoby05} and 700 newly discovered by 
\citet{reidparker05} ($\sim$ 200 confirmed by
\citealt{leisy05}); SMC 132 estimated and 101 discovered
\citet{jacoby05} ($\sim$ 70 confirmed by \citealt{leisy05}); M32 30
\citet{ciardullo89} (14 confirmed by \citealt{richer02}); NGC~205 35
\citet{corradi05} (13 confirmed by \citealt{richer02}); IC~10 16
\citet{magrini03}; NGC~6822 17 \citet{leisy} (6 confirmed, Leisy,
private communication); NGC~185 5 \citet{corradi05} (5 confirmed by
\citealt{richer02}); NGC~147 9 \citet{corradi05}; Sagittarius 3
\citet{zijlstra05} (confirmed); Fornax 1 \citet{danziger78}
(confirmed); Pegasus 1 \citet{jacoby}; Leo~A 1 \citet{magrini03}
(confirmed, Leisy, private communication); NGC~3109 18
\citet{prada05}; Sextans~B 5 \citet{M02} (confirmed by
\citealt{magrini05}); Sextans~A 1 \citet{magrini03} (confirmed by
\citealt{magrini05}).

Despite the large errors, a linear trend can be traced (the
continuous line in the upper panel of Figure~\ref{Fig-inter}),  
where the total stellar mass of the galaxies
is plotted versus the number of PNe, for all cases in which we could
estimate their population size corrected to a completeness limit of 4
magnitudes below the PNLF cutoff.  This number is extrapolated
from the empirical formula of the luminosity function \citep{jacoby89}
for the LG galaxies where the completeness limit in the search for PNe
was known, and where PNe above this limit were discovered.
From this plot we can
infer which galaxies had relatively strong star formation during
the past $\sim$1--8~Gyrs, i.e. in the period corresponding
to the formation of the PNe progenitors.  In particular, we can
compare the location of dwarf galaxies in Figure~\ref{Fig-inter} with
their SFHs reviewed by \citet{mateo}.  We find that the galaxies which
show little star formation during the past 1--8~Gyrs, i.e. the
galaxies with a relative star-formation rate $<$0.2 during that peroid
of time (cf. Figure~8 of \citealt{mateo}), generally lie in the
diagram below the continuous line (filled squares).  On the contrary,
those with strong intermediate-age star formation, namely
SFR$\geq$0.2, are located above the least squares fit line (filled
circles).We
note that galaxies which had a conspicuous star formation during the
past $\sim$1--8~Gyrs lie above the continuous line, with the only
exception being NGC~6822. It has a smaller number of PNe than
expected for its mass. It might be due to the determination of its 
stellar mass which is particularly difficult because of the large extent of its halo
\citep{weldrake}. Another aspect which might be
suggested from Figure~\ref{Fig-inter} is that all systems associated
with M~31 (M~32, NGC~205, NGC~147, IC~10, with exception of NGC~185)
and with the MW (SMC, LMC)
have enhanced 'recent' star formation, while several isolated systems
have lower rates (Sextans~A, Sextans~B, Leo~A, 
NGC~6822), suggesting that star formation might be
enhanced by interaction with the giant galaxies.

In the lower panel, 
the total stellar mass of the galaxies
is plotted versus the observed  number of candidate PNe. 
Again a linear trend can be traced, but we note that 
only a lower limit to the intermediate-age mass can be estimated 
from this plot because the number of PNe might be not complete. 
The location in this diagram of
IC~1613 and WLM, is in agreement with what we know about their star
formation histories, showing a  higher star formation during
intermediate ages in IC~1613 than in WLM.

PNe are therefore confirmed to be useful evolutionary age tracers of the 
intermediate-age population. The presence of PNe is enough evidence for 
an intermediate-age population 
\citep{apariciogallart94}, and, in addition, the relationship of their number to the 
host galaxy mass gives  information on the relative star formation history 
between different galaxies, where the same completeness in the search of PNe has been 
reached. These results agree with the those obtained by accurate SFH
analysis \citep{mateo}.

\section{Summary and Conclusions}

In this paper we presented the search for PNe in three dwarf irregular
galaxies belonging to the Local Group, or close to it: IC~1613, WLM,
and \gr8.  We discovered  two new candidate PNe in IC~1613, one in
WLM, and none in \gr8.  Their number and their absolute magnitude were
analyzed as a function of the SFH of the hosting galaxies.  The number
of PNe of these galaxies together with the number of PNe in the other
LG galaxies are used to estimate the mass of intermediate-age
population of each galaxy. This is compared with the SFR of the LG
dwarf galaxies from 1 to 8~Gry ago, finding that the number of PNe
agrees with the results obtained by accurate SFH
analysis and can therefore be used to constrain synthetic models of
SFHs.

\section*{Acknowledgments}

We are very grateful to the anonymous referee whose comments resulted in a significant 
improvement of the work reported in this paper. 

The INT data have been  made publically available through the Local Group 
Census programme (PI: N. A. Walton) of the Isaac Newton Group's
Wide Field Camera Survey Programme. This research has made use of the NASA/IPAC Extragalactic Database (NED) 
which is operated by the Jet Propulsion Laboratory, California Institute of Technology, under contract with the National 
Aeronautics and Space Administration,  
the APM Sky Catalogue and USNO-A2.0 Sky Catalogues, and the ESO Online Digitized
Sky Survey.  DM is supported by FONDAP Center for Astrophysics
15010003.
 
\label{lastpage}

\end{document}